\documentclass[preprint]{aastex}

\newcommand{\feh}{\hbox{$ [\mathrm{Fe}/\mathrm{H}]$}}

\newcommand{\vbump}{\hbox{$ \mathrm{M_{V,{\mathrm{b}}}}$}}
\newcommand{\rbump}{\hbox{$ \mathrm{R_{b}}$}}

\shorttitle{RGB Bump}
\shortauthors{Bjork \& Chaboyer}

\begin{document}
\title{Theoretical Uncertainties in Red Giant Branch Evolution: The
Red Giant Branch Bump}

\author{Stephan R.\ Bjork  and Brian Chaboyer}
\affil{Department of Physics and Astronomy, Dartmouth College,
6127 Wilder Lab, Hanover, NH 03755}
\email{Brian.Chaboyer@Dartmouth.edu}

\begin{abstract}

A Monte Carlo simulation exploring uncertainties in standard stellar
evolution theory on the red giant branch of metal-poor globular
clusters has been conducted. Confidence limits are derived on the
absolute $V$-band magnitude of the bump in the red giant branch
luminosity function (\vbump)  and the excess number of stars in the
bump, \rbump.  The analysis takes into account uncertainties in the
primordial helium abundance, abundance of alpha-capture elements,
radiative and conductive opacities, nuclear reaction rates, neutrino
energy losses, the treatments of diffusion and convection, the surface
boundary conditions, and color transformations.

The uncertainty in theoretical values for the red giant bump magnitude
varies with metallicity between $+0.13/-0.12$ mag at $\feh = -2.4$
and $+0.23/-0.21$ mag at $\feh = -1.0$. The dominant sources of
uncertainty are the abundance of the alpha-capture elements , the
mixing length, and the low-temperature opacities.  The theoretical values of
\vbump\ are in good agreement with observations.  The uncertainty in
the theoretical value of \rbump\ is $\pm 0.01$ at all metallicities
studied. The dominant sources of uncertainty are the abundance of the
alpha-capture elements, the mixing length, and the high-temperature
opacities. The median value of \rbump\ varies from 0.44 at $\feh =
-2.4$ to $0.50$ at $\feh = -1.0$.  These theoretical values for
\rbump\ are in agreement with observations.

\end{abstract}


\section{Introduction\label{intro}}

Globular clusters are made up of a very large number of stars with
varying mass but identical age, composition, and distance. This makes
them a rich and productive application of the theory of stellar
structure and evolution. Detailed stellar evolution calculations are
done numerically using computer programs which incorporate previously
calculated nuclear reaction rates and opacities, approximations to
complex phenomena such as convection, and assumptions about the
chemical composition. To compare theoretical models to observations,
moreover, requires converting physical quantities such as luminosity
and surface temperature to the observational system of magnitudes and
colors using empirical relations or the results of separate stellar
atmosphere models. The results of theoretical stellar evolution
calculations, therefore, depend upon a set of prior assumptions. To
assess the reliability of theoretical models of globular clusters
stars one must study how uncertainties in these assumptions of stellar
evolution theory propagate to the predictions of the theory.

In this work we are interested specifically in studying uncertainties
in stellar evolution theory along the red giant branch (RGB) of metal-poor
globular clusters. The RGB is the region in a
color-magnitude diagram comprising low-mass stars that have already
expended the hydrogen fuel at their center, and are burning hydrogen
in a spherical shell around an inert helium core. Stars in this stage
of evolution grow larger, brighter, and redder over time. An excellent
review of RGB evolution is found in \citet{sal02},
including a discussion of uncertainties in the
theoretical models and observational tests of the theory. Much of our
work was guided by this review.

After a low-mass star has consumed all the hydrogen at its center, the
envelope of the star expands and it moves across the color-magnitude
diagram from the main sequence region towards lower effective
temperature and redder colors. Stars in this stage are said to be on
the subgiant branch. Initially, hydrogen burning continues through a
thick shell covering a region of mass $\sim0.1M_\odot$, while outside
the hydrogen-burning shell the stellar envelope has the original
chemical composition with a convective region in the star's outer
layers. As the star progresses along the subgiant branch, the
hydrogen-burning shell narrows to a mass of $\sim0.001M_\odot$ while
the convective region grows steadily deeper. The convective region
eventually reaches material previously processed by the
hydrogen-burning core during the main sequence phase, and containing a
higher abundance of helium which immediately is mixed throughout the
convective region.

Eventually, the stellar luminosity begins to grow as the effective
temperature continues to drop, and the star is said to be at the base
of the RGB. The star progresses up the RGB
as the hydrogen-burning shell moves outward through the star leaving
behind an increasingly massive core of helium. The star's radius
continues to grow larger, its effective temperature lower, and its
luminosity brighter. The lower boundary of the convective region,
which reached a maximum depth near the base of the RGB,
recedes steadily.

A discontinuity in the chemical composition is left at the maximum
depth reached by the convective envelope, since convection mixes
hydrogen from the envelope into the partially depleted region left
behind by the hydrogen-burning core of the main sequence phase. When
the hydrogen-burning shell reaches this discontinuity, the sudden
increase in available fuel causes the stellar luminosity to drop
temporarily, and the star's evolution pauses before continuing its
progression up the RGB. This point is called the RGB bump.

There is a very strong correlation between stellar luminosity and the
mass of the helium core along the RGB. The luminosity
grows as the core mass grows, and since the rate at which mass is
added to the helium core is itself proportional to the luminosity,
this means that throughout red giant evolution the helium core mass
and the luminosity increase at an ever faster rate. The one exception
to this is at the RGB bump, where the sudden change in chemical
composition causes the growth in luminosity to pause temporarily,
while the helium core continues to gain mass.

The probability of observing a star in a given luminosity range on the
RGB is inversely proportional to the rate at which stars
evolve at that luminosity. In observed globular cluster
color-magnitude diagrams, therefore, we find that the number of
observed stars steadily decreases along the RGB, except
at the red giant bump. In fact, the differential luminosity function
of a globular cluster, showing the number $N$ of observed stars as a
function of magnitude, descends along the RGB as a nearly
straight line in the magnitude-$\log N$ plane. The slope of this line
indicates how the rate of evolution increases along the RGB.

Globular cluster luminosity functions provide an important test of the
accuracy of stellar evolution models. The number of stars observed as
a function of luminosity indicates the relative timescale of stellar
evolution, which in turn conveys information about the internal
chemical structure of stars. A great deal of attention has been paid,
for example, to the total number of stars on the RGB
compared to the main sequence turn-off region. Several authors 
\citep{bolte,van98,langer}
have found a discrepancy in this quantity between
observed luminosity functions for the globular clusters M5 and M30 and
the predictions of stellar evolution theory. \citet{van98}
take this discrepancy to suggest the presence of rapidly
rotating cores in red giants, while \citet{langer} take it
to indicate the possibility of a chemical mixing process deep in the
stellar interior.  

The magnitude of the RGB bump in observed globular
clusters serves to indicate the depth of the chemical discontinuity
left by the convective envelope at its point of deepest extent near
the base of the RGB. In one of the first extensive
studies of the red giant bump, \citet{fusi} compared
observational determinations of the bump magnitude in 11 globular
clusters to theoretical predictions. They found that, taken relative
to the horizontal branch magnitude, theoretical values of the bump
magnitude are higher than observed values by about 0.4 mag. However,
more recent studies using updated stellar evolution models and
improved observations \citep{cas3,zoccali,riello} 
have found no discrepancy between theory and observations.

\citet{cas3} have studied how uncertainties in the most
important individual stellar evolution parameters impact the magnitude
of the red giant bump. They look at the equation of state, mixing
length, mass loss along the RGB, opacities, and the $V$-band
bolometric correction. In addition, \citet{cas1} studied the impact of
element diffusion in detail. The effects of overshooting from the
convective envelope have been considered by \citet{alongi}.  This
paper presents a comprehensive study of \vbump\, incorporating all the
relevant uncertainties to firmly establish the agreement between
standard stellar models and observations of the red giant bump.

The magnitude of the RGB bump depends upon critically upon the maximum
depth of the convection zone.  In contrast, the enhancement in the
observed number of stars in the bump depends upon the size of the
chemical discontinuity. \citet{bono} introduced the \rbump\ parameter,
which measures the relative number of stars in the RGB bump.  
\citet{bono} found that \rbump\  was a robust
prediction of stellar evolution theory, as changes in the opacity,
equation of state, and nuclear cross sections changed the predicted
value of \rbump\ by a few percent.  Their work found fair agreement between
the observed \rbump\ values and those predicted by standard stellar
evolution theory.  More recently \citet{riello} used HST data to
determine \rbump\ in 54 Galactic globular clusters.  They found this
data was in good agreement with standard stellar evolution models.  

This paper presents a comprehensive analysis of how uncertainties in
the inputs into stellar evolution theory effect the predictions for
\vbump\ and \rbump.  Section \ref{method} provides an overall of the
Monte Carlo approach used in the paper, while \S \ref{uncertain}
presents a detailed analysis of the uncertainties in standard stellar
evolution theory.  The theoretical luminosity functions are presented
in \S \ref{theolf}.  Section \ref{rgbb} discusses the \vbump\ results,
while the \rbump\ results are presented in \S \ref{rbmp}.  A summary
of the main results is given in \S \ref{summ}.

\section{Method\label{method}}

Since it is not feasible to use analytic error propagation formulas
with the complex calculations of stellar evolution, one must use numerical
methods to study how uncertainty in the parameters of stellar
evolution theory propagate to the predictions of the theory. A simple
approach is to consider one parameter at a time, performing
calculations with several different values of the parameter to examine
how the predictions of the theory are affected \citep[e.g.][]{cast99}.
The effects involved are generally small, so
changes in the parameters produce a roughly linear response in results
of the calculations. However, this method does not describe how
uncertainties in many different parameters interact to produce a
combined uncertainty in the predictions of the theory. To investigate
this requires a larger set of calculations, in which all of the
significant parameters vary simultaneously.

We use a Monte Carlo approach to the problem. Describing the estimated
uncertainty in each parameter with a probability distribution, we run
a large number of independent stellar evolution calculations in which
the value of each parameter is drawn randomly from its corresponding
probability distribution. For specific quantitative predictions of the
theory, then, the results of all the runs can be combined into
histograms showing the most probable value and the distribution of
uncertainty around it. This method has been used previously in a
series of papers examining uncertainties in theoretical globular
cluster ages \citep{chab96,chab98,chab02,krauss}. The work here is essentially
an extension of this method to study theoretical uncertainties in RGB 
evolution.

Stellar evolution is strongly affected by the abundance of heavy
elements in a star, indicated by the relative iron abundance
[Fe/H]. Because the iron abundance (or metallicity) of a star can be
determined from observations, it is treated as a known quantity in our
simulation rather than as a source of theoretical uncertainty. We
therefore conduct our analysis at five different metallicity values
spanning a range typical of old galactic globular clusters. For each
set of randomly generated parameters in the Monte Carlo simulation, we
calculate an evolutionary sequence using [Fe/H] = $-$1.0, $-$1.3,
$-$1.6, $-$2.0, and $-$2.4. The results are analyzed separately for
each metallicity.

Our stellar evolution calculations are made using the Dartmouth
stellar evolution program \citep{chab01},
which is a descendant of the Yale Rotating Stellar Evolution Code
\citep{yrec}. To explore the evolution of metal-poor stars
along the RGB, we calculated evolutionary sequences for a
variety of masses and metallicities starting from previously
constructed zero-age main sequence models. Evolution was carried up to
the tip of the RGB, which we defined as the point where
energy generated by the triple-alpha reaction reached 1\% of the
star's total luminosity.  The equations of stellar structure were
solved to an accuracy of a few parts in $10^5$. With these tolerances,
the code requires about 6000 -- 7500 individual stellar models to
evolve the highest metallicity stars to the tip of the RGB, 
and for the lowest metallicity stars it requires 3500 -- 4000
models. Each stellar model comprises about 600 spherical shells on the
main sequence, and approximately 1500 shells along the RGB.
The greatest difficulty in the Monte Carlo approach is that it is
computationally very demanding. This study required 45,000
evolutionary sequences, roughly 230 billion stellar models, and
approximately 600 CPU days on a Beowulf cluster of 2.0Ghz Athalon
processors.

In all the stellar evolution calculations we use the standard DSEP
equation of state, which includes the Debye-H\"uckel correction of
\citet{clayton} for electrostatic interactions in the fully ionized
plasma. The DSEP equation of state in the fully ionized regime ($T >
10^6$~K) is essentially the relativistic ideal gas law with the
electrostatic interaction, radiation pressure, and electron degeneracy
effects included. In the partially ionized regime it is a Saha
equation including single ionization of hydrogen and metals, and
double ionization of helium. Although a more sophisticated equation of
state is now available in the OPAL tables of \citet{rogers}, \citet{chab95}
have found that using the OPAL tables in place of the
standard DSEP equation of state does not change the evolution of
metal-poor stars significantly. Based on their work, we consider the
equation of state to be an insignificant source of uncertainty in
metal-poor red giant evolution.

Similarly, all of our calculations use the standard treatments of
\citet{salpeter54} and \citet{grab} to evaluate plasma
screening effects on the nuclear reactions. More recent treatments of
plasma screening are available \citep[e.g.,][]{mitler}. However, \citet{cas2}
report that their numerical experiments found the new
formulations to have no impact on metal-poor red giant evolution. We
therefore do not consider plasma screening as a source of uncertainty.

The other important parameters of stellar evolution
theory---opacities, nuclear reaction rates, composition, surface
boundary conditions, neutrino cooling rates, and the treatments of
convection and diffusion---represent possible sources of uncertainty,
and so they are varied throughout the Monte Carlo simulation. The
treatment of these parameters is detailed in the next section.

\section{Sources of Uncertainty\label{uncertain}}

There are 18 parameters of stellar evolution theory that we consider
as possibly significant sources of uncertainty. Some of these, like
the primordial helium abundance, are simple numerical values that vary
through the Monte Carlo simulation. Others, though, like the opacities
and nuclear reaction rates, enter the stellar evolution calculations
as tables or formulas. These quantities are varied by multiplying the
table or formula by an overall coefficient, and then varying the
coefficient through the Monte Carlo simulation. In the case of
high-temperature opacities, we use different but correlated
multiplicative factors in different regions of the table. For the
bolometric corrections, which are logarithmic quantities, we add
rather than multiply by an overall uncertainty term.

During the Monte Carlo simulation, the value of each parameter is
drawn randomly from a probability density distribution that we choose
to reflect the estimated uncertainty in the parameter. In general, we
use Gaussian distributions to reflect uncertainties that are
statistical in nature, with a well-defined most probable value. For
systematic uncertainties we use uniform distributions in which the
probability density is constant within some range and zero outside of
it. For the surface boundary conditions there are two independent
formulations to choose from, and here we use a binary distribution in
which one or the other formulation is chosen with equal probability.

The 18 parameters varied in the Monte Carlo simulation are listed in
Table \ref{table1} along with the probability density distributions
used in the Monte Carlo simulation. In many cases, we adopt a
distribution from one of the previous papers in which Monte Carlo
simulations were used to investigate stellar evolution models at the
main sequence turn-off point and on the subgiant branch
\citep{chab96,chab98,chab02,krauss}. For others, we review the
literature directly to estimate an appropriate distribution, and these
are discussed in detail.

\noindent \textbf{Convective overshoot}. One of the weaknesses of the 
mixing length approximation used in standard stellar evolution models
is that it fails to properly describe the boundary between convective
and radiative regions in a star. Assuming no composition gradient,
this boundary is established by the Schwartzschild condition as the
point where the temperature gradient in the star is equal to the
adiabatic temperature gradient.  The mixing length approximation
assumes that all convective motion is confined to this region of
instability.

In reality, convective material reaches the boundary of the unstable
region with a non-zero velocity, and therefore tends to penetrate
across the Schwartzchild boundary into the stable radiative
region. This phenomenon is called convective overshoot, and it has
been demonstrated in hydrodynamical simulations of convection 
\citep[e.g.,][]{singh}. The extent to which overshooting
actually occurs in stars, however, is unclear.

Since the stars in our simulation do not have convective cores, we are
concerned only with overshooting at the base of the surface convection
zone. The stellar evolution calculations model convective overshooting
by assuming that the region in which the chemical composition is
homogenized by convective mixing extends some distance below the
Schwartzschild boundary into the stable radiative region. Temperature
gradients, however, are taken to be unaffected by overshooting. The
single parameter in this treatment is the depth by which convective
mixing reaches into the radiative region, and this distance is
expressed as a fraction of the pressure scale height at the
Schwartzschild boundary.

An upper limit on convective overshooting in metal-poor stars is set
by observed lithium abundances. Lithium breaks down in the high
temperatures of stellar interiors, beginning near the bottom of the
surface convection zone. A large amount of convective overshooting
would carry lithium more quickly from the surface to the interior, and
lithium depletion would occur at a faster rate. Overshooting depths
greater than about 0.2 $H_p$ would be inconsistent with the relatively
high lithium abundances observed in metal-poor stars.  
In our simulation, therefore, we
draw values for the overshooting depth from the uniform distribution
0.0 -- 0.2 $H_p.$

\noindent \textbf{Diffusion coefficients}. Models of stable
astrophysical plasmas predict that helium and heavy elements in a
star's radiative regions should settle toward the center of the star
over time, while hydrogen rises toward the surface. The extent to
which this actually occurs, however, is subject to considerable
uncertainty. There is clear evidence from helioseismology that element
diffusion occurs in the sun 
\citep{dals,basu}. Studies of surface abundances in
metal-poor stars, though, indicate that in these stars diffusion for
some reason does not occur in the outer layers \citep{chab01}.

Our stellar evolution calculations incorporate element diffusion using
the treatment of \citet{thoul}. Because diffusion is
not seen in the outer layers of metal-poor stars, we include a
modification introduced by \citet{chab01} which suppresses
diffusion near the surface. By comparing observed iron abundances in
globular clusters to the iron abundance predicted by their stellar
evolution models, \citet{chab01} estimate that whatever process
inhibits diffusion in metal-poor stars must act over an outer region
of at least 0.005 $M_\odot$. On the other hand, the fact that
diffusion is not found to be inhibited in the sun suggests that the
process inhibiting diffusion extends no lower than the bottom of the
solar surface convection zone, which has a mass of $0.02
M_\odot$. Therefore \citeauthor{chab01} conclude that the process
inhibiting diffusion likely acts over a surface region with mass
somewhere between 0.005 and 0.02 $M_\odot$.

To reflect this in our calculations we follow \citet{chab01}
in setting the diffusion coefficients to zero in the
outer 0.005 $M_\odot$ layer of the star. In the interior region
defined by $M_* - M(r) > 0.02\,M_\odot$ where $M_*$ is the star's
total mass, the diffusion coefficients are set to the standard values
of \citet{thoul}.  In the middle region
$0.005\,M_\odot < M_* - M(r) < 0.02\,M_\odot$, the coefficients are
ramped from zero to the standard values.

To reflect the uncertainty in this treatment of diffusion, we multiply
the diffusion coefficients overall by a factor drawn from the uniform
distribution 0.5 -- 1.3. This range is chosen based on an estimated
30\% uncertainty in the theoretical calculations of diffusion
velocities \citep[see e.g.][]{chab96}, but the lower end of the
range is extended to account for the fact that several physical
processes in the interior of a star might slow the rate of diffusion,
while no such processes would increase it. 

\noindent \textbf{Triple-alpha reaction rate}. The point at which 
helium ignition occurs in a star's evolution depends on the rate of
the helium-burning triple-alpha reaction, in which three helium nuclei
fuse to form a single nuclei of $^{12}\mathrm{C}$.  Our calculations
use the rate of \citet{cf88}, with an uncertainty
estimated at 15\% \citep[e.g.][]{cast99}. We
therefore multiply the standard \citet{cf88} rate by a factor
drawn from the Gaussian distribution $1.00 \pm 0.15$.

\noindent \textbf{Plasma neutrino cooling rate}. Neutrino cooling in
the stellar core becomes an important effect in the evolution of stars
on the upper RGB. Energy lost through the emission of
neutrinos tends to slow the rise in temperature as the helium core
grows, allowing the core to reach a higher mass before helium burning
begins. Significant energy loss can occur through several different
neutrino processes in a stellar plasma; for the helium core of stars
on the upper RGB the most important is the plasma
process, in which a plasmon (a quantized electromagnetic oscillation
in the plasma) spontaneously decays into a neutrino-antineutrino pair
\citep{itoh96}.

We use the plasma neutrino cooling rates of \citet{haft}, 
who have calculated numerical rates based on the treatment of
electromagnetic dispersion relations in plasmas provided by \citet{braaten},
 and have also derived an analytic approximation to the
numerical results for use in stellar evolution calculations. They find
their analytic approximation accurate to within 4\% in regimes where
the plasma process is dominant and within 5\% everywhere.

Published neutrino cooling rates are generally calculated under the
standard assumption that neutrinos have no magnetic dipole moment. A
non-zero magnetic moment for neutrinos would tend to increase the
plasma neutrino emission rates and could impact stellar evolution
significantly. In fact, \citet{raff90,raff92} and \citet{cast93}
have used stellar evolution calculations
incorporating a non-zero neutrino magnetic moment to constrain the
magnetic moment through comparison to globular cluster observations.

Although the existence of a non-zero magnetic moment could
significantly impact stellar evolution at the tip of the RGB, 
this possibility is considered at the moment to be outside of
standard physics, and so we do not incorporate it into our
analysis. Instead, we consider the error in the \citet{haft}
analytic approximation formula to be the dominant source of
uncertainty in the neutrino cooling rates, and to reflect this in the
Monte Carlo simulation we multiply the \citeauthor{haft} formula
by a factor drawn from the Gaussian distribution $1.00 \pm 0.05$.

\noindent \textbf{Conductive opacities}. In the deep interior of RGB 
stars, thermal conduction by electrons serves as an
important means of energy transfer. The thermal conductivity under
these conditions therefore serves to establish the temperature
gradient in red giant interiors and determine when the core becomes
hot enough to ignite helium.

Conductive opacities are one of the most significant sources of
uncertainty on the upper RGB, so a close examination of
our current understanding in this area is appropriate. There are
essentially two treatments of conductive opacity available for use in
red giant models: the tabulated values of \citet{hubbard} and
the more recent calculations of \citet{itoh83}. However, as
\citet{cat96} point out, the \citet{itoh83} calculations were
made for conditions characteristic of white dwarfs and neutron stars;
red giant cores do not fall within their range of
validity. Specifically, the \citeauthor{itoh83} results are presented as an
analytic fitting formula in terms of the parameter \[\Gamma =
\frac{Z^2 e^2}{r k_b T}\] which characterizes the strength of the
electrostatic interaction between ions in a plasma. Here $e$ is the
electron charge, $Z$ is the atomic number, $k_B$ is the Boltzmann
constant, and $r = [3/(4 \pi n_i)]^{1/3}$ is the ion-sphere radius
with $n_i$ the number density of ions. The domain where $2 \lesssim
\Gamma \lesssim 171$ represents matter in the liquid metal phase,
while $\Gamma \lesssim 2$ corresponds more typically to a Boltzmann
gas. The fitting formula of \citet{itoh83} intended for application to
dense matter in the liquid metal phase, is valid only in the range $2
\le \Gamma \le 160$, while in red giant cores $\Gamma$ is considerably
lower. Applying the results of \citet{itoh83} to RGB stars
therefore requires extrapolating the fitting formula from liquid metal
conditions to Boltzmann gas conditions, and it is not clear how much
error this extrapolation introduces. Because of this, \citet{cat96}
conclude that the \citet{hubbard} values are to be preferred at
present.

The treatment of conductive opacity in our stellar evolution
calculations is based on the \citet{hubbard} tables, along with the
relativistic extension provided by \citet{canuto} for higher
densities. Specifically, we employ a set of analytic fitting formulas
from \citet{sweig}; for $\log \rho \le 5.8$, we use Sweigart's fit
to the \citeauthor{hubbard} tables, while for $\log \rho \ge 6.0$ we use
Sweigart's fit to the Canuto relativistic opacities. For $5.8 < \log
\rho < 6.0$ we use a ramp between the non-relativistic and
relativistic formulas, also provided by \citet{sweig}.  For conditions
appropriate to red giant stars, we find that Sweigart's formula are
accurate to about 10\% when compared to the \citet{hubbard}
tabulations.  

We estimated the uncertainty in conductive opacities by directly
comparing values from the \citet{itoh83} analytic formula to the
tabulated values of \citet{hubbard}. \citet{cat96} have
plotted both these sets of opacities for conditions typical of red
giant cores. Their plot shows that here the \citeauthor{itoh83} values run
$\sim\! 10$ -- $30\%$ lower than the \citeauthor{hubbard} values. Since
the \citeauthor{itoh83} fitting formula was not designed to be accurate in
this regime, though, it is possible that this comparison could
exaggerate the difference between the two treatments. To check this,
we also compared the two sets of opacities in a density-temperature
region where both treatments are valid.

We make the comparison at densities typical of red giant cores
($\rho\!\sim\!10^5$ -- $10^6 \,\mathrm{g/cm^3}$). The acceptable
temperature range is then bounded by two restrictions in the
\citet{itoh83}  analysis. On the upper end there is the restriction that
$\Gamma > 2$. On the lower end is the requirement that a
high-temperature classical limit apply to the treatment of the ionic
system. This is expressed by the condition $y \ll 1$ where the
parameter $y$ is defined \[y \equiv \frac{\hbar^2 k_F^2}{2Mk_BT}\]
with $k_F$ being the Fermi wavenumber of the electrons and $M$ the
mass of an ion. \citet{mitake} found that the
high-temperature classical limit is adequate as long as $y < 0.01$. In
the range $0.01 < y < 0.1$, though, their results show that a quantum
correction not included in the \citet{itoh83} analysis tends to
reduce the conductive opacities by up to 25\%.

In Figure \ref{copacfig} we have plotted the conductive opacities at
densities of $10^{5.0}$ and $10^{5.5}\, \mathrm{g/cm^3}$ for
temperatures where $\Gamma > 2$ and $y < 0.1$, and we show the
temperatures corresponding to $y =0.01$. The agreement between the
different treatments is better at the higher temperatures, where the
classical treatment of ions by \citeauthor{itoh83} is most valid and also
where the temperatures approach those typically found in red giant
cores ($T \sim 10^{7.5}$ -- $10^{8.0}$ K). In the region where $y <
0.01$ the treatments agree to within 25\%, with the \citeauthor{itoh83}
values running 5 -- 25\% below the \citet{hubbard} values.

Taking all of this into consideration, we estimate that current values
for the conductive opacity in red giant cores are uncertain by about
20\% at the 1-$\sigma$ level. We therefore multiply our standard
values (obtained as described above from Sweigart's fit to the \citeauthor{hubbard}
tables and the relativistic Canuto treatment) by a factor
drawn from the Gaussian distribution $1.00 \pm .20$.

\indent \textbf{Bolometric corrections}. In order to compare our 
stellar evolution models to observations, it
is necessary to transform the physical quantities used in stellar
evolution calculations to those quantities measured by
observers. Specifically, we must transform the theoretically
calculated bolometric luminosity, effective temperature, and surface
gravity to predict the observational magnitudes for specific pass
bands. Generally, this transformation is made using color tables based
on theoretical treatments of stellar atmospheres, possibly calibrated
by empirical data. Purely empirical relationships between magnitude
and effective temperature are also available based on effective
temperatures measured for nearby stars, but these relationships are
valid only for a restricted range of metallicities and evolutionary
stages.

The transformations used in our analysis are based on the color tables
constructed for the Revised Yale Isochrones \citep{ryi}. 
These transformations are an empirical recalibration of the
theoretical colors and bolometric corrections of \citet{van85}
and \citet{kurucz}. In this study we will examine the $V$-band
magnitude of the RGB bump, so we must estimate the
uncertainty in the $V$-band bolometric corrections.  \citet{weiss}
compare the $V$-band bolometric corrections $(BC_V)$ from
several theoretical, empirical, and semi-empirical sources and find
that with a consistent choice for the solar $BC_V$, all the sources
agree to within 0.05 mag along a typical globular cluster isochrone.
Our own analysis of different $(BC_V)$ from different sources confirm
this result.  Therefore, for $BC_V$, we draw the uncertainty term from
the uniform distribution $\pm$0.05 mag.

\section{Theoretical Luminosity Functions\label{theolf}}

To explore uncertainties in \vbump\ and \rbump, we conducted a
Monte Carlo simulation with 1120 independent sets of randomly chosen
stellar evolution parameters, from each set generating luminosity
functions at several different metallicities, ages, and assumed
initial mass functions. 

Generating luminosity functions requires a series of theoretical
evolutionary tracks calculated for stars over a range of masses. In
this work we use the nine masses $M$ = 0.55, 0.63, 0.70, 0.75, 0.80,
0.85, 0.90, 0.95, and 1.00 $M_\odot$. Each evolutionary track is
calculated from the zero-age main sequence to the tip of the RGB.
The range of masses was chosen to allow reliable luminosity
functions for ages between 10 and 20 Gigayears.

From a set of evolutionary tracks at appropriate masses one can
calculate theoretical isochrones at a variety of ages. We use a
modified version of the isochrone generating program used to construct
the Revised Yale Isochrones \citep{ryi}. The program uses the
method of equivalent evolutionary points, locating in each
evolutionary track a set of points defined in terms of the central
helium abundance (on the main sequence) or the helium core mass (on
the RGB) and interpolating among them to generate the
isochrones. From these isochrones, $V$-band luminosity functions are
calculated.  Uncertainty in the bolometric correction is incorporated
at a later stage of the analysis. Luminosity functions are normalized
to 1000 stars on the zero-age main sequence and use a bin size of 0.04
mag.

Generating a luminosity function requires an initial mass function
(IMF) describing the relative number of stars as a function of mass
created during the formation of the cluster. The IMF is generally
assumed to have the form of a simple declining power law, $\xi(M)
\propto M^{-\alpha}$. In the classic study, \citet{salpeter} found that
observational data suggested an initial mass function with exponent
$\alpha$~=~2.35; more recent studies, though, find evidence that the
initial mass function for globular clusters varies from cluster to
cluster and that metal-poor clusters in general have less steep IMFs
than metal-rich clusters.

Because stars on the RGB represent a narrow range of
masses, though, the luminosity function in the red giant region is
less affected by the IMF than are the main sequence and subgiant
branch. In particular, we expect that \vbump\ and \rbump\ will be 
unaffected by the IMF.  This is tested by 
generating luminosity functions using the \citet{salpeter} value for the IMF
exponent as well as values from extremes of the reasonable range:
$\alpha$~=~0.00, 2.35, and~4.00.

\begin{figure}
\plotone{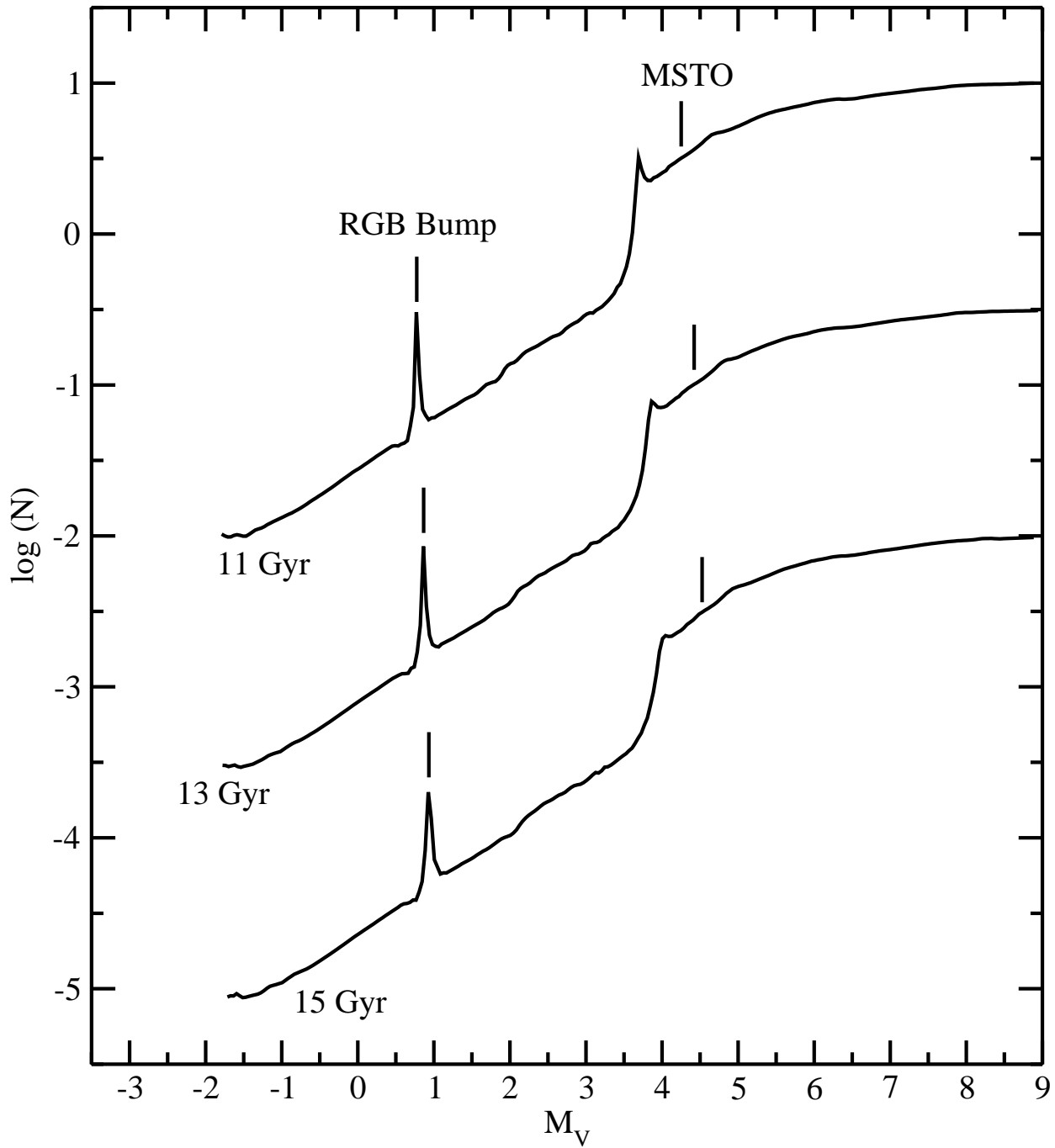}
\caption{Luminosity functions calculated using a value for each
stellar evolution parameter from the center of its uncertainty range,
$\feh = -1.0$ 
and using the Salpeter IMF $\alpha$~=~2.35. 
The location of the RGB bump  and the main sequence turn-off point are
indicated. \label{lffig}}
\end{figure}

\begin{figure}
\epsscale{0.8}
\plotone{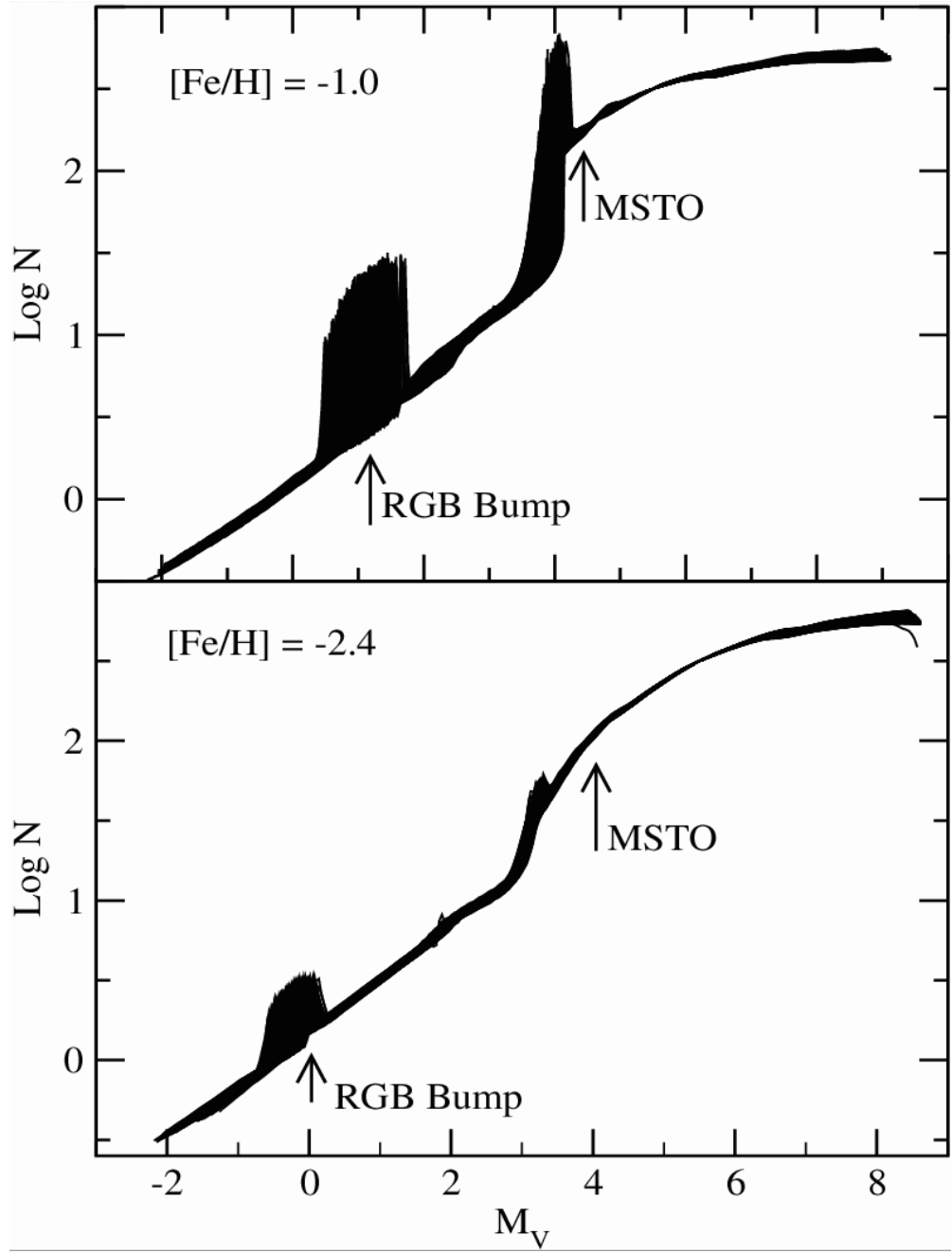}
\caption{Monte Carlo luminosity functions calculated 
using the Salpeter IMF $\alpha$~=~2.35, an age of 13 Gyr and $\feh =
-1.0$ (upper panel) and $\feh = -2.4$ (lower panel).
\label{mclffig}}
\end{figure}

Sample luminosity functions are shown in Figure \ref{lffig}.  These
luminosity functions were calculated using values for the stellar
evolution parameters from the center of the uncertainty ranges
discussed in Section \ref{uncertain}. Luminosity functions are shown
for a single metallicity at ages of 11, 13, and 15 Gigayears using the
Salpeter value for the IMF exponent, $\alpha$~=~2.35.  The entire set
of Monte Carlo luminosity functions (1140) for a Salpeter IMF, an age
of 13 Gyr and $\feh = -2.4$ and $\feh = -1.0$ are shown in Figure
\ref{mclffig}.  In making this figure, the luminosity functions have
been normalized so that they all have the same number of stars at $M_V
= 5.5$, which is approximately 1.5 mag below the turn-off.

\section{The Magnitude of the Giant Branch Bump\label{rgbb}}

To precisely locate the peak of the RGB bump in our calculated
luminosity functions, we find the point in the
luminosity function which corresponds to the most populated bin in the
bump region, and then calculate the parabola passing through that
point and the adjacent points on either side. The vertex of the
parabola is taken as the peak of the bump, and the magnitude of the
vertex as the bump magnitude.

As discussed in Section \ref{uncertain}, the uncertainties in the
\citet{ryi} bolometric corrections taken into consideration by 
 adding to the bump magnitude a
term drawn randomly from the uniform range $0.00\pm0.05$ mag. For each
of the 1120 Monte Carlo realizations at a given age and metallicity we
generate ten such uncertainty terms, so that 1120 independent
luminosity functions are used to produce 11,200 realizations of the
bump magnitude. These realizations are independent in the
bolometric correction, though not in other stellar evolution
parameters. The increase in the number of points does not
significantly improve the numerical accuracy of the Monte Carlo
simulation, but it does provide smoother histograms.

Studies of the RGB bump \citep[e.g.][]{fusi} often focus on
the difference in magnitude between the bump and the zero-age
horizontal branch level, $\Delta V^\mathrm{bump}_\mathrm{HB} =
V_\mathrm{bump} - V_\mathrm{HB}$. Observationally, this quantity is
more reliable than $V_\mathrm{bump}$ since it avoids uncertainties in
the distance modulus and in the calibration of observational data. The
theoretical evaluation of $\Delta V^\mathrm{bump}_\mathrm{HB}$,
however, is subject to uncertainties in the luminosity of the
horizontal branch. In order to confine our analysis to uncertainties
on the RGB, we will consider the absolute magnitude $M_V$ of the RGB
bump rather than $\Delta V^\mathrm{bump}_\mathrm{HB}$.

\begin{figure}
\plotone{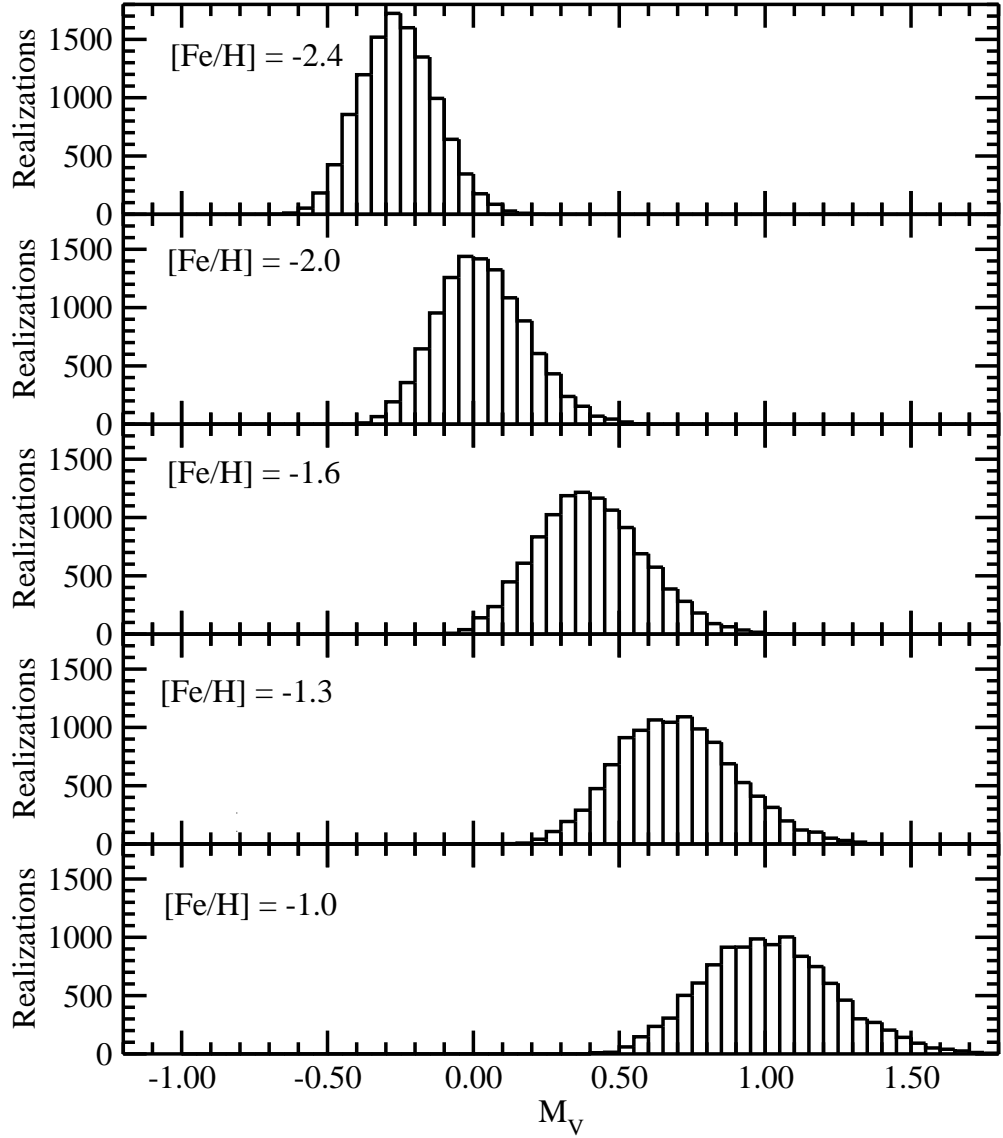}
\caption{The distribution of RGB bump magnitudes at age 13 Gyr.\label{bmphist}}
\end{figure}

Table \ref{table2} shows the median values of the bump magnitude obtained in
the Monte Carlo simulation, along with 68\% and 95\% confidence
limits. Both the median values and the confidence limits depend
strongly on the cluster metallicity. In contrast, the choice of
cluster age serves merely to shift the distributions of bump magnitude
without affecting their shape, so the same confidence limits apply at
all ages. In order to show the shape of the bump magnitude
distributions, we plot the results for age 13 Gigayears as histograms
in Figure \ref{bmphist}.

Both Table \ref{table2} and Figure \ref{bmphist} show that the
uncertainty in theoretical determinations of the bump magnitude
increases with metallicity. At [Fe/H]~=~$-$1.0 the distribution of
bump magnitudes obtained in the Monte Carlo simulation is about 80\%
wider than at [Fe/H]~=~$-$2.4. At all metallicities the distributions
show a slight asymmetry, falling off more gradually toward dimmer
magnitudes.

The effect of cluster age on the bump magnitude is essentially the
same at all metallicities. An increase of 1 Gigayear in age
corresponds to a shift towards dimmer bump magnitudes of 0.03 -- 0.04
mag, not a large effect compared to the width of the magnitude
distributions. Even at the lowest metallicity, where uncertainties in
stellar evolution theory leave the tightest distribution of bump
magnitudes, the shift in magnitude between age 11~Gyr and age 15~Gyr
is only 60\% of the distribution width at the 68\% confidence level.

The effect of metallicity on the bump magnitude is much more
significant. Changing the cluster metallicity by $\pm 0.1$ to  0.2 dex
leads to a shift in the bump magnitude comparable to the 68\%
theoretical confidence limits in Table \ref{table2}.  Since
observational measurements of [Fe/H] are uncertain by about $\pm$0.15
dex, this means that when we compare observational and theoretical
determinations of the bump magnitude the uncertainties in stellar
evolution theory are about as important as observational uncertainties
in the metallicity scale.

\begin{figure}
\plotone{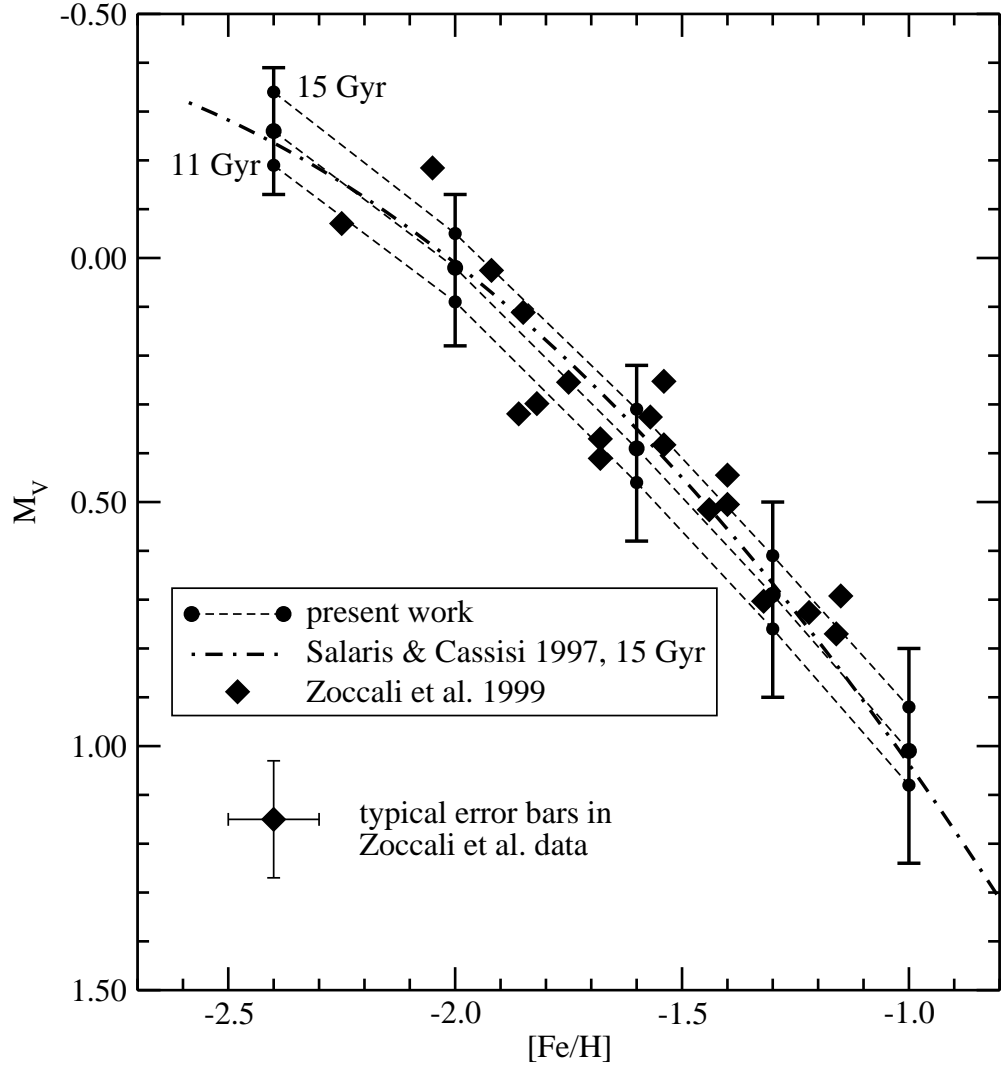}
\caption{$M_V$-band magnitude of the RGB bump as a function of
metallicity for the ages of 11, 13 and 15 Gyr.  The error bars
represent correlated 68\% confidence limits in our results, are shown
only for age 13 Gyr.  The same confidence limits apply at all
ages. Also shown is the theoretical relation of \citet{cas3}.  The
observational results of \citet{zoccali} assuming the globular cluster
distance scale of \citet{krauss}.  Each observational point has an
associated uncertainty of approximately $\pm$0.10 dex in [Fe/H] and
$\pm$0.12 mag in $M_V$.
\label{bmpfeh}}
\end{figure}

Figure \ref{bmpfeh} shows the bump magnitude as a function of
metallicity, compared to other theoretical and observational
results. Error bars are shown representing the 68\% confidence limits
at age 13 Gyr, but we omit the error bars at ages 11 and 15 Gyr for
clarity. The theoretical relation of \citet{cas3} for clusters of age
15 Gigayears is also shown, as well as observational results for 19
galactic globular clusters drawn from \citet{zoccali}. 
\citeauthor{cas3} present their results in terms of the global
metallicity [M/H], and they assume a different value than we do for
the solar heavy element abundance $Z_\odot$. To make their treatment
of metallicity agree with ours, we shift their results to the [Fe/H]
scale assuming [$\alpha$/Fe]~=~0.45 (the median value in our Monte
Carlo simulation) and $Z_\odot$~=~0.018. The \citeauthor{cas3}
relation is somewhat lower than our results for age 15 Gyr, especially
at the highest and lowest metallicities. The difference, however, is
well within the theoretical uncertainties.

The \citet{zoccali} data points show the $M_V$-band magnitude of
the red giant bump in 19 Galactic globular clusters observed with the
\emph{Hubble Space Telescope}. We use the \citet{zinn}
metallicities for each cluster, and convert from apparent to absolute
magnitude using the RR Lyrae calibration of \citet{krauss}:
$M_V(\mathrm{RR}) = 0.46 + 0.23\, ([\mathrm{Fe/H}]+1.9)$. The
uncertainty in the absolute magnitude of the bump is dominated by an
uncertainty of $\pm$0.12 mag in the RR Lyrae calibration, while the
uncertainty in globular cluster metallicities is about $\pm$0.15 dex. 

The agreement between the \citet{zoccali} observations and our
theoretical models is excellent. \citeauthor{zoccali} themselves found a
good agreement between their observational values of $\Delta
V^\mathrm{bump} _\mathrm{HB}$ and the models of \citet{cas3}
assuming an alpha-capture overabundance [$\alpha$/Fe]~=~0.30
for [Fe/H] $<$ $-$1.0. Our results show, similarly, a good agreement
between observations and theoretical values of $V_\mathrm{bump}$
calculated with a median alpha-capture overabundance
[$\alpha$/Fe]~=~0.45. Moreover, the results of our Monte Carlo
uncertainty analysis show that differences between theoretical and
observed values of the bump magnitude lie within the theoretical
uncertainties in stellar evolution as well as within the observational
uncertainties in the globular cluster distance modulus and the
metallicity scale.   It is clear that there is currently no discrepancy
between the $V$-band bump magnitude as observed in Galactic globular
clusters and as calculated in standard stellar evolution models.

To analyze the individual influence of each continuously varying
stellar evolution parameter on \vbump, all 1120 Monte Carlo
realizations are plotted  on a graph of \vbump\  versus parameter
value.  The dependence of \vbump\ on a given  parameter value
is then characterized with a straight line fit. Specifically, the
parameter range is divided into 20 bins containing 56 Monte Carlo
realizations each, and for each bin the median value of \vbump\  and the 68\% confidence limits are determined. A straight
line is fit to the median points of each bin using the simple
least-squares method with each bin weighted equally. Straight lines
are also fit to the upper and lower 68\% confidence points in each
bin. Table \ref{table3} presents the slope of the linear
dependence for the most significant stellar evolution parameters as
well as the total change in bump magnitude as the parameter varies
across its range of uncertainty, either between the endpoints of a
uniform distribution or between the 68\% confidence limits of a
Gaussian distribution. The impact of surface boundary conditions is
explored by comparing the Monte Carlo distribution of bump magnitudes
obtained using the Eddington $T(\tau)$ relation to that using the
\citet{krish} relation, and Table \ref{table3} 
records the difference
between the median bump magnitudes obtained with each of the two
treatments.

The three most significant sources of uncertainty in the bump
magnitude are the alpha-capture abundance, the mixing length, and the
low-temperature opacities. All three of these parameters have a
stronger influence at higher metallicities, which explains why the
overall uncertainty in the bump magnitude increases with
metallicity. Other significant sources of uncertainty are the surface
boundary conditions, convective overshoot, the
$\mathrm{^{13}C} + p \rightarrow \mathrm{^{14}N} + \gamma$ reaction
rate, and the helium diffusion coefficients, 
The impact of these parameters does not vary significantly
with [Fe/H], and so the values given in Table \ref{table3} are averaged across
all five metallicities. All of the other stellar evolution parameters
explored in the Monte Carlo simulation are negligible, impacting the
bump magnitude at the level of 0.05 mag or less.

\citet{cas3} and \citet{cas1}
have already studied the impact of a few of these stellar
evolution parameters on theoretical bump magnitudes. Exploring the
influence of mixing length, \citeauthor{cas3} find a dependence
$\Delta V_\mathrm{bump}/\Delta \mathrm{ml} \approx -0.27$ mag with
$\Delta$ml in units of $H_p$ and with the metallicity unspecified, a
result that is in rough agreement with our more detailed
analysis. \citet{cas1} consider the impact of
diffusion on the bump luminosity, and find that including helium and
heavy element diffusion according to the \citet{thoul}
coefficients lowers the bump luminosity by $\Delta V_\mathrm{bump}
\approx 0.07$ mag compared to models with no diffusion. The impact of
diffusion in our Monte Carlo simulation is comparable.
 
It is interesting that convective overshoot represents only a
secondary source of uncertainty in the bump magnitude. After one of
the first studies of the RGB bump \citep{fusi} found a
substantial discrepancy of 0.4 mag between observational and
theoretical determinations of $\Delta V^\mathrm{bump}_\mathrm{hb}$,
\citet{alongi} suggested that this discrepancy could be
resolved by including convective overshoot with extension 0.7 $H_p$ in
the theoretical models. More recent comparisons using newer
observations and theoretical models \citep{cas3,zoccali}
have found that the discrepancy is resolved even without
convective overshoot. In addition, though, our results show that with
an upper limit of 0.2 $H_p$ set by observed lithium abundances in
metal-poor stars, the question of whether to include convective
overshoot in stellar evolution models is in fact of only secondary
importance, much less significant than the alpha-capture abundance,
mixing length, and low-temperature opacities.

Finally, we consider the confidence limits in the
bump magnitude obtained when the most significant parameters are held
fixed at the center of their uncertainty range, indicating the
accuracy that theoretical bump magnitudes would have if one of the
important stellar evolution parameters were known precisely. Table \ref{table4}
shows the 68\% confidence limits for fixed values of the alpha-capture
overabundance, mixing length, and low-temperature opacities. The
median bump magnitudes in Table \ref{table2} for each metallicity and age still
apply. Particularly important are the confidence limits obtained for
fixed values of [$\alpha$/Fe], both because this parameter is the
greatest source of uncertainty and because it is in principle an
observable quantity. The confidence limits in Table \ref{table4} indicate that
if the alpha-capture abundance could be precisely determined for a
given cluster, the uncertainty in the theoretical bump magnitude for
that cluster would be reduced by about 35\%. Improved spectroscopic
observations of globular clusters allowing direct measurements of the
alpha-capture abundance will therefore be the most important step
toward improving the accuracy of theoretical bump magnitudes.

\section{Number Counts of Stars in the Bump\label{rbmp}}

The relative enhancement of the number of stars in the RGB bump can be
measured using the \rbump\ parameter introduced by \citet{bono}.  It
measures the relative evolutionary timescale for stars in the bump
region to stars on the RGB which have not yet reached the bump region.
The \rbump\ parameter was measured in each of the MC luminosity
functions, using the definition of \citet{bono}: the ratio between the
star counts in the bump region $\mathrm{V_{bump}}\pm 0.4\,$mag to the
star counts fainter on the RGB: $\mathrm{V_{bump}} + 0.5 <
\mathrm{V_{bump}} < \mathrm{V_{bump}} + 1.5$.  In agreement with
\citet{bono} we find that \rbump\ is a robust prediction of stellar
evolution theory, with little uncertainty.  The \rbump\ parameter has
a small dependence on \feh\ and age, and the results from the MC
simulation are summarized in Table \ref{table5}.  It was found that,
at a given \feh\ and age, the $1\,\sigma$ uncertainty in the
predication of \rbump\ is $\pm 0.01$.  Parameters which effect \rbump\
the most are, in order of significance, $[\alpha/\mathrm{Fe}]$, mixing
length, the high temperature opacities, helium diffusion and the low
temperature opacities.  The relatively modest effect of each of these
parameters on \rbump\ is summarized in Table \ref{table6}.

\begin{figure}
\plotone{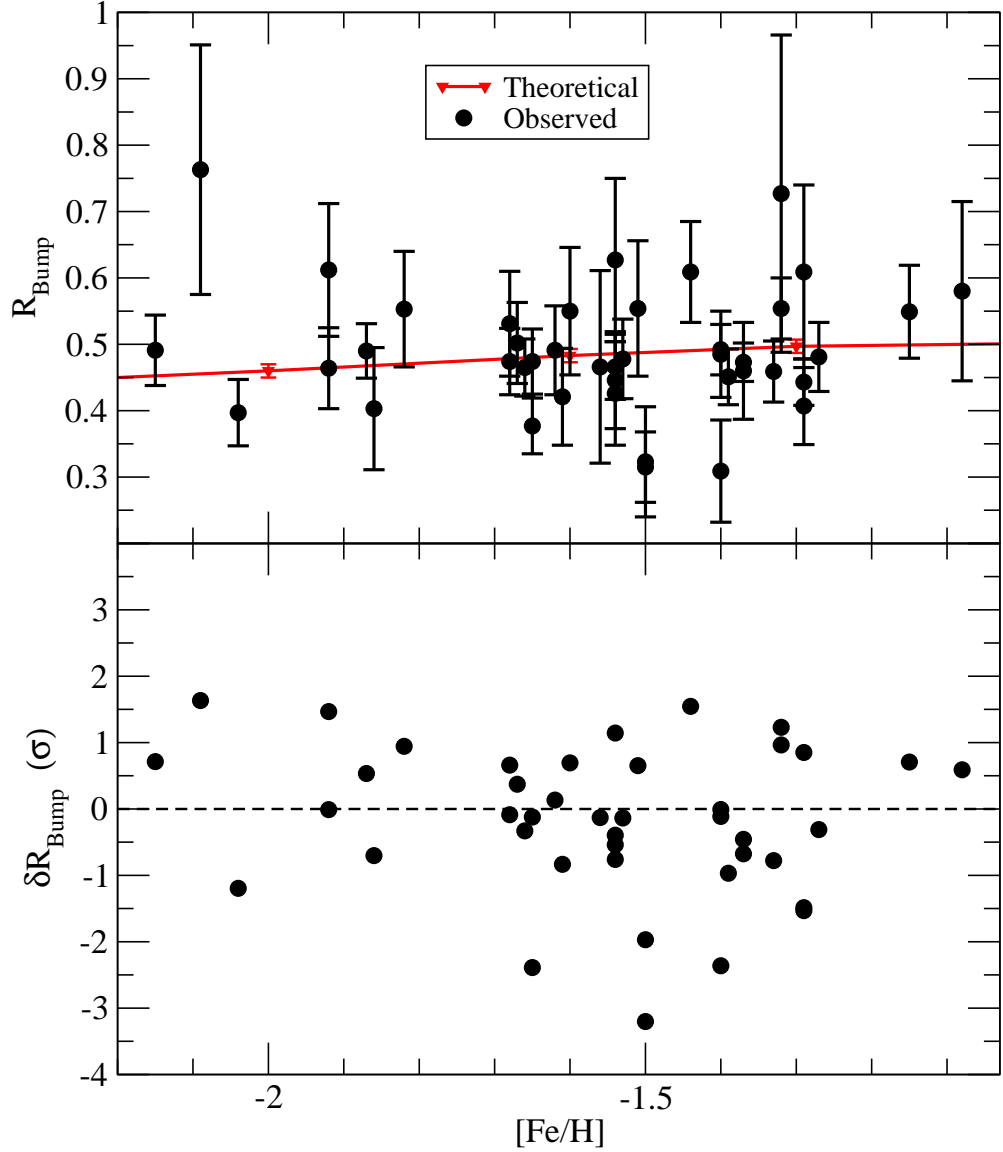}
\caption{ Comparison between \rbump\ predicated by theory and
obserations (see text for references) as a function of the
\citet{zinn} metallicity.  The bottom panel is the residuals between
the theoretical relation and the observevations.  These residuals have
been normalized by dividing the actual residual by the error in each
of individual observations added in quadrature with the error in the 
theoretical predication.
\label{rbmpfeh}}
\end{figure}

Comparison between observed and predicated values of \rbump\ serves as
an excellent test of standard stellar evolution theory.
\citet{riello} used HST data to determine \rbump\ in 54 Galactic
globular clusters, of which 40 are metal-poor ($\feh \le -1.0$) and
can be compared to our MC results. To this data set, we have added
additional determinations of \rbump\ from studies which presented high
quality luminosity functions of the globular clusters M3 \citep{rood},
M5 \citep{san96}, M10 \citep{pol05}, M12 \citep{har04}. Each of these
observational studies presented luminosity functions based upon
observations of $>10,000$ stars in a given cluster.  The comparison
between these data and the MC models is shown in Figure \ref{rbmpfeh},
assuming the \citet{zinn} metallicity scale.  For this metallicity
scale our theoretical predication has a reduced $\chi^2 = 1.23$ when
compared to the observations.  This implies that our models provide a
reasonable fit to the observations.  Formally, this value of a reduced
$\chi^2$ implies that the models have a 14\% probability of correctly
describing the data.  If the \citet{car97} metallicity scale is used,
then the reduced $\chi^2 = 1.34$ (6.5\% propability) and the models
provide an acceptable fit to the data.

The observed luminosity function of M10 was found to disagree with the
predications of standard stellar evolution theory \citep{pol05}, due to 
the fact that there are too many stars on the RGB compared to the
main-sequence.  From the \citet{pol05} luminosity function, M10 has 
$\rbump = 0.55\pm 0.10$, compared to a predicted value of 
$\rbump = 0.49\pm 0.01$ on the \citet{car97} metallicity scale and 
$\rbump = 0.48\pm 0.01$ on the \citet{zinn} metallicity scale.   Thus,
the excess number of stars on the main-sequence has not lead to 
an anomalous number of stars in the bump region on the RGB.

\section{Summary\label{summ}}

We use a large Monte Carlo simulation to investigate theoretical
uncertainties in the RGB luminosity function, specifically in the
$V$-band magnitude of the red giant bump and excess 
number of stars in the bump.  We find excellent
agreement between theoretical and observational values for the bump
magnitude, with the uncertainty in theoretical values comparable to
the scatter in observational values. Metallicity has a very
significant effect on the bump magnitude, while the cluster age
impacts the bump magnitude at a level lower than the theoretical
uncertainties. The most important sources of uncertainty in the
predication of the bump magnitude are the alpha-capture overabundance
[$\alpha$/Fe], the mixing length, low-temperature opacities, and the
treatment of surface boundary conditions.  Theoretical uncertainties in 
stellar evolution models have little impact on the excess number of 
stars in the bump region. This is a robust prediction of
standard stellar evolution is found to be in reasonable agreement 
with observations.   

\acknowledgments
Research supported in part by a NSF CAREER grant 0094231  to Brian Chaboyer.
Dr.\ Brian Chaboyer is a Cottrell Scholar of the Research Corporation.

\clearpage

\begin{figure}
\plotone{f1.eps}
\caption{The fractional difference between the conductive opacities of
\citet{itoh83} and \citet{hubbard}. \label{copacfig}}
\end{figure}

\clearpage 

\clearpage 

\clearpage

\clearpage

\clearpage 

\clearpage

\begin{deluxetable}{llcc}
\tablecaption{Stellar Evolution Parameters in the Monte Carlo
Simulation.
\label{table1}}
\tablewidth{0pt}
\tablehead{
\colhead{Parameter } & 
\colhead{Standard }   & 
\colhead{Distribution } &
\colhead{Type} 
}
\startdata

He mass fraction (Y) & n/a & $0.245 - 0.250$ & unif. \\ 
$[\alpha/\mathrm{Fe}]$ & n/a & $0.2-0.7$ dex & unif.\\
Mixing length & n/a & $1.85 \pm 0.25 \: H_p$ & gaus.\\ 
Convective overshoot & n/a & $0.0-0.2 \: H_p$ & unif.\\
\hline
Atmospheric $T(\tau)$ & grey or \citet{krish} & 50/50 & binary\\
\hline
Low-T opacities & \citet{opacalex} & $0.7 - 1.3$  & unif.\\ 
High-T opacities & \citet{opal} & $1.00 \pm 2\%$ \tiny{($T \ge 10^7$K)} & gaus.\\ 
&& $0.98 \pm 4\%$ \tiny{($T \le 10^6$K)} &  \\ 
Diffusion coefficients & \citet{thoul} & $0.5 - 1.3$ & unif. \\ 
$p+p \rightarrow  {^2 \mathrm{H}} + e^+ + \nu_e$ & \citet{nucrate} & $1 \pm 0.2\%$ & gaus. \\
&& $1\:^{+0.14\%}_{-0.09\%} \: ^{+2.0\%}_{-1.2\%}$ & unif.\\ 
$\mathrm{{^3He} + {^3He} \rightarrow {^4He}} + 2p$ & \citet{nucrate} & $1 \pm 6\%$ & gaus. \\
$\mathrm{{^3He} + {^4He} \rightarrow {^7Be}} + \gamma$ & \citet{nucrate} & $1 \pm 3.2\%$ & gaus. \\
$\mathrm{^{12}C} + p \rightarrow \mathrm{^{13}N} + \gamma$ & \citet{nucrate} & $1 \pm 15\%$ & gaus. \\
$\mathrm{^{13}C} + p \rightarrow \mathrm{^{14}N} + \gamma$ & \citet{nucrate} & $1 \pm 15\%$ & gaus. \\
$\mathrm{^{14}N} + p \rightarrow \mathrm{^{15}O} + \gamma$ & \citet{nucrate} & $1 \pm 12\%$ & gaus. \\
$\mathrm{^{16}O} + p \rightarrow \mathrm{^{17}F} + \gamma$ & \citet{nucrate} & $1 \pm 16\%$ & gaus. \\
Triple-$\alpha$ reaction rate & \citet{cf88} & $1 \pm 15\%$ & gaus. \\ 
Neutrino cooling rate & \citet{haft} & $1 \pm 5\%$ & gaus. \\ 
Conductive opacities & \citet{hubbard} & $1 \pm 20\%$ & gaus. \\ 
$BC_V$ & \citet{ryi}& $\pm 0.05$ mag & unif.\\

\enddata

\tablecomments{Parameters in the lower portion of the table represent
multiplicative or additive factors modifying standard tables or
formulas. The atmospheric $T(\tau)$ relation is handled as a discrete
choice between two independent formulas.}
\end{deluxetable}

\begin{deluxetable}{crrrrrccc}
\tabletypesize{\scriptsize}
\tablecaption{Absolute V Magnitude of the Red Giant Bump
\label{table2}}
\tablewidth{0pt}
\tablehead{
\colhead{}& 
\multicolumn{5}{c}{Median $M_V$}&  & 
\multicolumn{2}{c}{Confidence limits}\\
\cline{2-6} \cline{8-9}\\
\colhead{[Fe/H]} & 
\colhead{11 Gyr }   & 
\colhead{12 Gyr }   & 
\colhead{13 Gyr }   & 
\colhead{14 Gyr }   & 
\colhead{15 Gyr }   & 
\colhead{~~}& 
\colhead{68\%} &
\colhead{95\%} 
}
\startdata

$-2.4$& $-0.34$&  $-0.30$&  $-0.26$&  $-0.22$&  $-0.19$&&  $+0.13 /-  0.13$& ~ $+0.27  /-0.23$\\
$-2.0$& $-0.05$&  $-0.01$&  $ 0.02$&  $ 0.06$&  $ 0.09$&&  $+0.16 /-  0.15$& ~ $+0.32  /-0.27$\\
$-1.6$& $ 0.31$&  $ 0.35$&  $ 0.39$&  $ 0.43$&  $ 0.46$&&  $+0.19 /-  0.17$& ~ $+0.38  /-0.32$\\
$-1.3$& $ 0.61$&  $ 0.65$&  $ 0.69$&  $ 0.73$&  $ 0.76$&&  $+0.21 /-  0.19$& ~ $+0.42  /-0.35$\\
$-1.0$& $ 0.92$&  $ 0.97$&  $ 1.01$&  $ 1.04$&  $ 1.08$&&  $+0.23 /-  0.21$& ~ $+0.47  /-0.39$\\
\enddata
\end{deluxetable}

\begin{deluxetable}{lccrc}
\tablecaption{Effect of Significant Stellar Evolution Parameters on 
\vbump 
\label{table3}}
\tablewidth{0pt}
\tablehead{
\colhead{Parameter ($X$) } & 
\colhead{Range }   & 
\colhead{[Fe/H] }   & 
\colhead{{$\frac{\Delta \vbump}{\Delta X}$} \tablenotemark{a} }    & 
\colhead{$\Delta \vbump$ \tablenotemark{b}}   
}
\startdata
[$\alpha$/Fe]         & $0.2-0.7$ dex 
&  $-2.4$ & $ 0.53~~$ & $0.21$\\
&& $-2.0$ & $ 0.65~~$ & $0.26$\\
&& $-1.6$ & $ 0.81~~$ & $0.33$\\
&& $-1.3$ & $ 0.88~~$ & $0.36$\\
&& $-1.0$ & $ 0.96~~$ & $0.39$\\[1ex]

Mixing length  &$1.85\pm0.25\,H_p$
&  $-2.4$ & $ -0.23~~$ & $0.15$\\
&& $-2.0$ & $ -0.29~~$ & $0.18$\\
&& $-1.6$ & $ -0.35~~$ & $0.22$\\
&& $-1.3$ & $ -0.42~~$ & $0.27$\\
&& $-1.0$ & $ -0.50~~$ & $0.32$\\[1ex]

Low-T opacities  &$0.7-1.3$
&  $-2.4$ & $ 0.16~~$ & $0.08$\\
&& $-2.0$ & $ 0.20~~$ & $0.10$\\
&& $-1.6$ & $ 0.24~~$ & $0.12$\\
&& $-1.3$ & $ 0.30~~$ & $0.14$\\
&& $-1.0$ & $ 0.36~~$ & $0.17$\\[1ex]

Convective overshoot  &$0.0-0.2\, H_p$& ~avg. & $ 0.47~~$ & $0.07$\\[.3ex]

Atmospheric $T(\tau)$  &Eddington / & ~avg. & ---~~~ & $0.07$\\[-.3ex]
                       &Krishna-Swamy &&&\\[.3 ex]

$\mathrm{^{13}C} + p \rightarrow \mathrm{^{14}N} + \gamma$ &$1\pm0.15$
& ~avg. & $ 0.24~~$ & $0.07$\\[.3ex]

He diffusion coef.  &$0.5-1.3$& ~avg. & $ 0.09~~$ & $0.06$\\[.3 ex]

\enddata

\tablenotetext{a}{The slope of the \vbump-parameter
dependence.} 
\tablenotetext{b}{Amount by which the bump
magnitude changes as the parameter varies across its 80\% uncertainty
range (see text for discussion).}
\end{deluxetable}

\begin{deluxetable}{lcc}
\tablecaption{
Confidence Limits on \vbump\ for a Fixed Individual Parameter
\label{table4}}
\tablewidth{0pt}
\tablehead{
\colhead{Parameter ($X$) } & 
\colhead{[Fe/H] }   & 
\colhead{68\% c.l.}   
}
\startdata
$[\alpha/\mathrm{Fe}]$  
                & $-2.4$&$+ 0.10/-   0.08$\\
                & $-2.0$&$+ 0.12/-   0.09$\\
                & $-1.6$&$+ 0.14/-   0.11$\\
                & $-1.3$&$+ 0.15/-   0.12$\\
                & $-1.0$&$+ 0.18/-   0.13$\\[1ex]
 Mixing length 
                & $-2.4$&$+ 0.12/-   0.10$\\
                & $-2.0$&$+ 0.15/-   0.12$\\
                & $-1.6$&$+ 0.18/-   0.14$\\
                & $-1.3$&$+ 0.19/-   0.15$\\
                & $-1.0$&$+ 0.20/-   0.17$\\[1ex]
 Low-T opacities  
                & $-2.4$&$+ 0.12/-   0.11$\\
                & $-2.0$&$+ 0.15/-   0.14$\\
                & $-1.6$&$+ 0.18/-   0.17$\\
                & $-1.3$&$+ 0.20/-   0.19$\\
                & $-1.0$&$+ 0.22/-   0.21$\\
\enddata
\end{deluxetable}

\begin{deluxetable}{crrrrrcc}
\tablecaption{Median \rbump\ Values  
\label{table5}}
\tablewidth{0pt}
\tablehead{
\colhead{[Fe/H]} & 
\colhead{Average}   &
\colhead{11 Gyr }   & 
\colhead{12 Gyr }   & 
\colhead{13 Gyr }   & 
\colhead{14 Gyr }   & 
\colhead{15 Gyr }   
}
\startdata

$-2.4$& $0.44$& $ 0.45$&  $ 0.45$&  $ 0.44$&  $ 0.43$&  $ 0.43$\\
$-2.0$& $0.46$& $ 0.47$&  $ 0.47$&  $ 0.46$&  $ 0.45$&  $ 0.45$\\
$-1.6$& $0.48$& $ 0.50$&  $ 0.49$&  $ 0.48$&  $ 0.48$&  $ 0.47$\\
$-1.3$& $0.50$& $ 0.51$&  $ 0.50$&  $ 0.50$&  $ 0.49$&  $ 0.49$\\
$-1.0$& $0.50$& $ 0.51$&  $ 0.50$&  $ 0.50$&  $ 0.50$&  $ 0.50$\\

\enddata
\tablecomments{In all cases the 68\% confidence level  is $\pm 0.01$ and
 the 95\% confidence level is $\pm 0.02$.}
\end{deluxetable}

\begin{deluxetable}{lccrc}
\tablecaption{Impact of Significant Stellar Evolution
Parameters on \rbump. 
\label{table6}}
\tablewidth{0pt}
\tablehead{
\colhead{Parameter ($X$) } & 
\colhead{Range }   & 
\colhead{[Fe/H] }   & 
\colhead{ {$\frac{\Delta \rbump}{\Delta X}$} \tablenotemark{a}  }   & 
\colhead{$\Delta \rbump$ \tablenotemark{b}}   
}
\startdata
[$\alpha$/Fe]         & $0.2-0.7$ dex 
&  $-2.4$ & $  0.032~~$ & $  0.013$\\
&& $-2.0$ & $  0.041~~$ & $  0.017$\\
&& $-1.6$ & $  0.043~~$ & $  0.018$\\
&& $-1.3$ & $  0.009~~$ & $  0.004$\\
&& $-1.0$ & $ -0.028~~$ & $  0.011$\\[1ex]

Mixing length  &$1.85\pm0.25\,H_p$ 
& avg. & $ 0.015~~$ & $ 0.010$\\[1ex]

High-T opacities  &$1\pm 0.02$
&  $-1.3$ to $-2.4$ & $ 0.12~~$ & $0.006$\\[1ex]

He diffusion  &$0.5-1.3$
&  $-1.3$ to $-2.4$ & $ 0.008~~$ & $0.005$\\[1ex]

Low-T opacities  &$0.7 - 1.3$
&  avg.& $ -0.009~~$ & $ 0.004$\\[1ex]

\enddata

\tablenotetext{a}{Slope of the \rbump-parameter
dependence.} 
\tablenotetext{b}{Amount by which \rbump\ changes
as the parameter varies across its 80\% uncertainty range.}
\end{deluxetable}

\end{document}